# Pressure induced 3D strain in 2D Graphene

Nathan Dasenbrock-Gammon,[1] Sachith Dissanayake,[2] and Ranga Dias*[1,2]

[1]*Department of Physics and Astronomy,*
*University of Rochester, Rochester, NY 14627, USA.*
[2]*Department of Mechanical Engineering,*
*University of Rochester, Rochester, NY 14627, USA.*

(Dated: July 28, 2022)

# Abstract

Two-dimensional (2D) materials such as graphene offer a variety of outstanding properties for a wide range of applications. Their transport properties in particular present a rich field of study. However, the studies of transport properties of graphene under pressure are mostly limited to ∼1 GPa, largely due to the technical challenges and difficulties of placing graphene inside a diamond anvil cell (DAC) and maintaining good electrical contacts under pressure. We developed a novel technique allowing for direct measurements of the transport properties of high quality chemical vapor deposition (CVD) monolayer graphene under pressures. Combined Raman spectroscopic and direct resistivity measurements on pure monolayer graphene up to 40 GPa shows an effective out of plane stiffness of $c_{33}=0.26\pm_{.09}^{.11}$ GPa, and observe relatively constant resistances with pressure, suggesting high pressure as a useful technique for producing large biaxial strains within graphene.

Since graphene was first isolated in 2004 [1], it has quickly become the subject of intensive research efforts. As the prototypical two-dimensional material, graphene offers a variety of outstanding properties for both theoretical and experimental studies. Owing to its exceptional electronic properties, extensive efforts have been made to study the transport properties of graphene under a variety of conditions - temperature [2], layering and orientation [3], stress and strain [4], etc. [5, 6]. Particularly exciting are the recent discoveries of superconductivity in magic angle graphene [3]. These new results promise a surge of new research in the field. In conjunction with angle, pressure provides another variable to tune the exciting behavior of these stacked graphene structures [7, 8]. Preliminary to studying the role of pressure on stacked graphene, the properties of monolayer graphene must first be characterized, yet, to the authors knowledge, these studies do not exist.

Superconducting behavior has been predicted in monolayer graphene [9], but experimental observation has proven challenging. Conventional superconductivity is largely driven by two factors: the density of states at the Fermi level and the electron-phonon coupling strength [10]. Monolayer graphene falls short on both of these fronts. First, graphene is a semi metal with a Dirac cone band structure and vanishing density of states at the Fermi surface. This drawback can be easily solved through doping, shifting the Fermi surface away from the Diarc point. The second drawback is not so easily rectified, though large dopant concentrations have been suggested to increase electron-phonon coupling [9], with some promising experimental results [11, 12]. Alternatively, large biaxial strains have been predicted to dramatically enhance the electron-phonon coupling of graphene [13, 14], with the goal of inducing conventional superconductivity.

The purpose of this work is to develop a novel technique to study the transport properties of monolayer graphene under pressure, utilizing high pressure diamond anvil cells, which will enhance our fundamental understanding of how graphene behaves under pressure. Such a study is essential for the development of new graphene-based electronic and optoelectronic devices. Despite extensive efforts to study graphene, work on graphene behavior under high pressure (greater than 1 GPa) remains limited. Diamond anvil cells (DACs) provide a system in which large biaxial strain, through pressure, can be applied to graphene. A number of vibrational studies have been performed on monolayer graphene within DACs up to pressures of 40 GPa [15–20]. A review of such studies can be found in [18]. There have also been a number of studies on stacked graphene or graphene heterostructures under pressure [7, 8, 21–23]. Regardless of these studies, work on the transport properties of monolayer graphene under pressure remains scarce. To remedy this, we have developed a novel technique allowing direct measurement of graphene resistance under pressure within a DAC and performed resistance measurements along with Raman spectroscopy measurements up to 40 GPa.

Commercially available Graphenea Easy Transfer product was used as the graphene source in this study. The product consists of a sheet of CVD graphene on a polymer substrate that is covered with a sacrificial poly(methyl/methacrylate) (PPMA) film. The transfer process is illustrated in Fig. 1, which follows the manufacturer's recommended transfer procedure adapted for use within a DAC. First, the Easy Transfer sheet is placed in a container of distilled water, causing the graphene/PPMA layers to release from the polymer substrate (Fig. 1a). This leaves a layer of graphene and PPMA floating on top of the water. One side of the DAC is brought up under the water, catching the graphene/PPMA on the diamond tip (Fig. 1b). After the sample is left to air dry, the graphene/PPMA is trimmed back to the culet area of the diamond by scraping the excess away. To promote graphene adhesion to the diamond, the DAC with diamond is held on a hot plate at $\sim$ 150 °C for 1 hour, which is followed being placed in a $\sim$75 torr vacuum for at least 16 hours. The last step of the transfer is to dissolve away the PPMA using warm (30-50 °C) acetone and isopropanol baths for one hour each, leaving only the monolayer graphene film on the diamond. Fig. 1e shows a schematic drawing of the graphene transferred to the diamond culet.

Verification of a successful transfer is performed via Raman spectroscopy of the diamond tip

[24]. A characteristic Raman spectrum is shown in Fig. 1f. An appealing aspect of this transfer technique is that it can be readily incorporated with measurements techniques with DAC, such as conductivity measurements. A state-of-the-art conductivity setup, adopting techniques described reference [25], is prepared on one side of the DAC. Tantalum electrodes ~150 nm thick are sputtered onto the diamond culet of the other side of the DAC using a custom fabricated shadow mask. To aid the adhesion of the electrodes to the diamond, several nanometers of chromium are first e-beam deposited before sputtering the tantalum. Following electrode deposition, the shadow mask is removed, which makes the DAC ready for graphene transfer. A complete experimental setup is depicted in Fig. 2, along with a schematic representation. A complete description of the conductivity setup and the electrode sputtering is provided in the supplementary methods.

Due to its two-dimensional nature, it is inherently difficult to make good electrical contact between graphene and three-dimensional metal electrodes [26], resulting in large contact resistances. These contact resistances, along with variation in graphene synthesis techniques, result in varied reported values of graphene sheet resistance [27]; typically in the order of several hundred ohms. This is in strong agreement with the four probe resistances measured here within the DAC before pressure is applied. Upon application of pressure, our observed resistances are slightly higher.

Raman scattering in graphene is well described by standard solid state procedures [28]. Monolayer graphene has three main Raman features, G band ~1580 cm$^{-1}$, D band ~1350 cm$^{-1}$, and 2D band ~2700 cm$^{-1}$. Due to diamond having Raman active modes in similar regions as graphene, many high pressure works utilizing diamond anvil cells focus on the G band. Similarly, in this work the D and 2D bands were obscured by Raman activity of the diamond anvils, and thus only the G band was measured with pressure - even if both the G and 2D bands were used at ambient pressure to verify the successful transfer of graphene. To translate these Raman spectroscopic measurements into mechanical information about graphene, we use a model presented by Sun *et al.* [20]. By ascribing a Morse potential to the in plane carbon-carbon stretching, they derive the G mode Raman frequency as a function of pressure:

$$\omega(P) = \frac{1}{\pi c}\sqrt{\frac{E_0 \beta^2 e^{\frac{\beta r_0 a_{33} P}{c_{11}^{2D}+c_{12}^{2D}}}(2e^{\frac{\beta r_0 a_{33} P}{c_{11}^{2D}+c_{12}^{2D}}}-1)}{m}}, \quad (1)$$

where $\beta$, $E_0$, and $r_0$ are the depth, width, and unstrained carbon-carbon distance, respectively, of the Morse potential, $a_{33}$ the interlayer spacing, and $c_{11}^{2D}$ and $c_{12}^{2D}$ are the 2D in plane elastic constants. Sun *et al.* applied this model to graphene by including an out of plane stiffness $c_{33}$ and

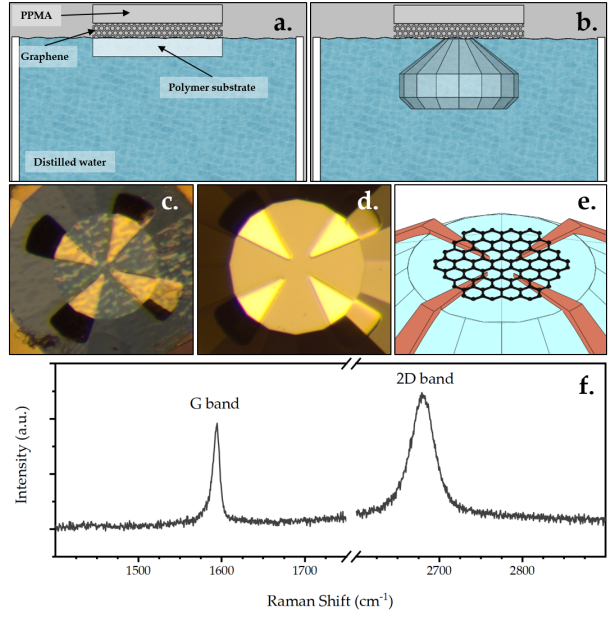

FIG. 1. a. The Graphenea monolayer graphene product consists of three layers. The monolayer graphene is situated between a polymer substrate and a PPMA top layer. The transfer process begins by bringing the Graphenea product into a container of water, releasing the graphene and PPMA layers, leaving them floating on the surface of the water. b. The graphene and PPMA layers are then caught by bringing the diamond up from under the water. Note that for clarity only the diamond was included in this rendering, the DAC assembly is not shown. c. A photograph of a diamond with the graphene and PPMA layers. To promote adhesion, the diamond/graphene is annealed at 150 °C for one hour before being held under moderate vacuum for at least 16 hours. d. The polymer layer is dissolved away through one hour baths in acetone and isopropanol, leaving just the graphene layer behind. Shown is a photograph of the diamond after PPMA has been dissolved away. e. Schematic image showing the graphene layer after being transferred onto the diamond. f. Verification of successful graphene transfer is performed via spectroscopic analysis. The G and D Raman modes are clearly visible, verifying a successful graphene transfer.

its shift with pressure $c_{33}$ through modifying $a_{33}$ in the above equation:

$$a_{33} = a_{33_0}\left[1 + \left(\frac{c'_{33}P}{c_{33}}\right)\right]^{-1/c'_{33}}. \tag{2}$$

The experimental data were fitted to the second derivative of this model up to ~2 GPa by fixing all parameters to that of graphite except $c_{33}$, and reported a value of $c_{33} = 1.4 \pm 295$ GPa. We expanded on their work, and performed Raman spectroscopic measurements up to 40 GPa.

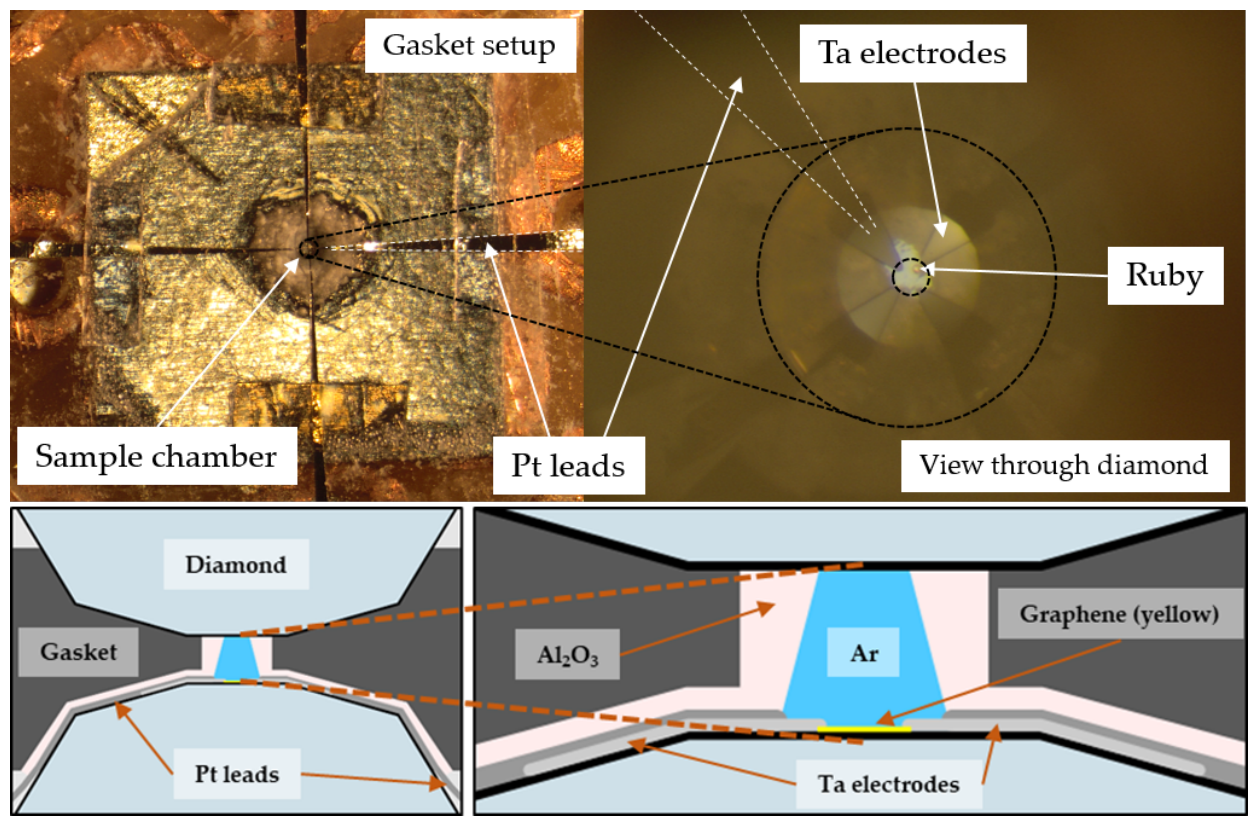

FIG. 2. Photographs and schematic views of conductivity setups used alongside the graphene transfer method. Top Left: The gasket is covered with insulating tape and the pre-indent packed with insulating $Al_2O_3$. Platinum leads are placed into the pre-indent. Top Right: A hole is made in the $Al_2O_3$ creating a sample chamber. Pressure determination is performed through the fluorescence of a coloaded ruby ball and the diamond Raman edge. The platinum leads make contact with sputtered tantalum electrodes over which the graphene was deposited. Finally, the sample chamber is high pressure gas loaded with an argon or neon pressure transmitting medium. Bottom left: Schematic view of the experimental setup used for resistance measurements with an argon pressure medium. Bottom right: A zoom-in on the sample chamber of the experimental setup used for resistance measurements with an argon pressure medium.

Raman measurements are summarized in Fig 3. Three separate experimental runs were performed. First, an argon pressure medium was gas loaded to 8 GPa, released to below 1 GPa, before being pressed to pass 20 GPa and subsequently released again. Two experimental runs were performed using a neon pressure medium. One was gas loaded to 2 GPa and pressed to 40 GPa before anvil failure, and a second was gas loaded to 9 GPa, pressed to 20 GPa, and then released. Slight differences can be seen between argon and neon as pressure media as shown in

$$\omega(P) = \frac{1}{\pi c} \sqrt{k \frac{E_0 \beta^2 e^{\frac{\beta r_0 a_{33} P}{c_{11}^{2D}+c_{12}^{2D}}} (2e^{\frac{\beta r_0 a_{33} P}{c_{11}^{2D}+c_{12}^{2D}}} - 1)}{m}} \qquad (3)$$

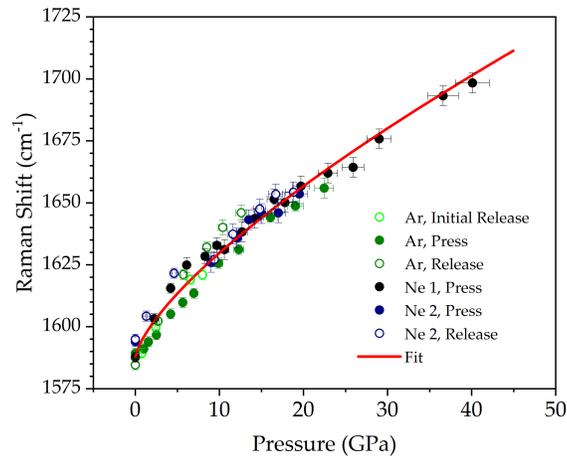

FIG. 3. The G band position as a function of pressure for several experimental runs. Slight differences can be seen between neon and argon pressure media. Uncertainty in peak comes from peak fitting around surface defect peaks from the diamond anvils, and pressure uncertainty is 5%. The solid red line shows the least squares fit of the model presented by [20] to all experimental runs.

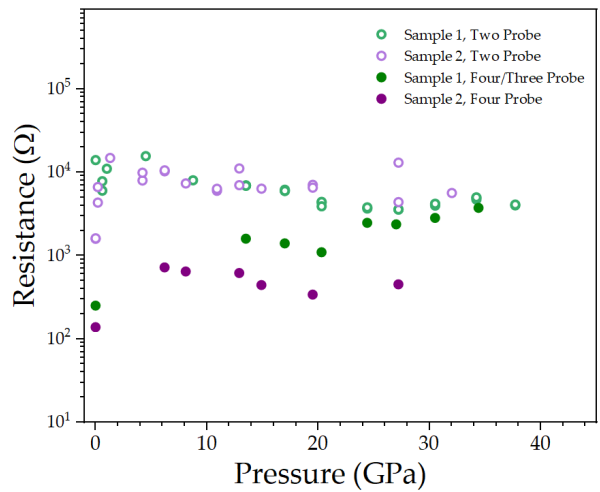

FIG. 4. Measured resistances of monolayer graphene with pressure using an argon pressure medium. Four/pseudofour probe measurements from two separate runs are shown. Resistances were observed to be relatively constant with pressure. Note that the jump in run 3 between 20 and 25 GPa corresponds to the loss of an electrode and switching between four and pseudofour probe resistance measurements.

Fig 3, in line with past work reported in reference [16]. Additionally, a mild hysteresis can be observed in the pressing vs releasing data. However, to obtain an estimate for the out of plane stiffness we combine all measured Raman shifts. We follow a similar approach to Sun *et al.* , taking the Morse potential parameters from reference [29] ($E_0$ = 6.13 eV, $\beta$ = 1.85 Å, and $r_0$ = 1.43 Å), obtained through ab initio calculations, and experimental values of $c_{11}^{2D}$ = 1109 GPa and $c_{12}^{2D}$ = 139 GPa [30]. As presented, this model drastically underestimates the zero pressure Raman frequency of graphene, yielding a value of $\sim$1380 cm$^{-1}$. Sun *et al.* prescribe this discrepancy to an underestimation of $E_0$, and in reference [31], this discrepancy is rectified by introducing a scaling factor in the model:

In this study we take $k = (\omega_0/\omega(0))^2$, where $\omega_0$ is the known zero pressure frequency for the graphene G band [24], and use $c_{33}$ and $c_{33}$ as fitting parameters. In the above approach we fix the model to match the zero pressure frequency for graphene, and extract information on graphenes out of plane stiffness from how the G band frequency changes under pressure. A least squares fit to $c_{33}$ and $c_{33}$ yields $c_{33}$=0.26$\pm_{.09}^{.11}$ GPa and $c_{33}$=3.44$\pm_{.29}^{.35}$. Our analysis significantly lowers the uncertainty on the previous measurement of $c_{33}$=1.4$\pm$295 GPa [20].

Studies on the transport properties of graphene under pressure are limited. Such studies either measure stacked graphene or other graphene heterostructures [7, 8, 21–23] and work on the transport properties of monolayer graphene under pressure remains scarce. We have performed four/pseudofour probe resistance measurements on graphene up to 38 GPa using an argon/neon pressure medium. These results are summarized in Fig 4. In contrast to Ke *et al.* [23] who observed a dramatic increase in resistance consistent with an opening bandgap in trilayer graphene, we observe a relatively constant resistance in monolayer graphene with only a slight increase at the start of the application of pressure. However, this contrast is not surprising, as with monolayer graphene there is no possibility for interlayer van der Waals coupling, which was the cause of their observed bandgap opening [32–34]. Strain can also result in an opening bandgap, among other things, and an entire field of straintronics exists [13]. It is known that a significant, ($\gtrsim$ 20%), uniaxial strain can lead to an opening bandgap in monolayer graphene [35–38]. However, by using a pressure medium the predominant strain within a DAC should be biaxial. Biaxial strain does not open a gap, as it preserves the honeycomb symmetry of graphene [13, 36], which explains why relatively flat resistances were observed with pressure, at least up to $\sim 40$ GPa.

In summary, we have developed a new technique allowing direct measurement of graphene resistance under pressure within a DAC and performed spectroscopic and resistivity experiments on monolayer graphene into previously unexplored pressure ranges. Resistivity measurements suggest that graphene resistance is relatively unchanged under the application of pressure up to 40 GPa, consistent with the pressure applying biaxial strain on the graphene. We extract an out of plane stiffness of $c_{33} = 0.26 \pm_{.09}^{.11}$ GPa, lowering the uncertainty on the previous measurement of $c_{33} = 1.4 \pm 295$ GPa [20]. These measurements lay the foundation for using DACs to generate the biaxial strain required to induce superconductivity in monolayer graphene. Further work to characterize the strains involved at these high pressures is desired, and the technique as presented here will be vital in exploring monolayer graphene in this regime of high pressure and strain.

This work was supported by National Science Foundation, Grant No. DMR-1809649.

# References

[20] Y. W. Sun, W. Liu, I. Hernandez, J. Gonzalez, F. Rodriguez, D. J. Dunstan, and C. J. Humphreys, 3d strain in 2d materials: To what extent is monolayer graphene graphite?, Physical Review Letters **123**, 10.1103/PhysRevLett.123.135501 (2019).

[21] F. Cellini, F. Lavini, T. Cao, W. de Heer, C. Berger, A. Bongiorno, and E. Riedo, Epitaxial two-layer graphene under pressure: Diamene stiffer than diamond, FlatChem **10**, 8 (2018).

[22] M. Yankowitz, K. Watanabe, T. Taniguchi, P. San-Jose, and B. J. Leroy, Article pressure-induced commensurate stacking of graphene on boron nitride, Nature Communications **7**, 10.1038/ncomms13168 (2016).

[23] F. Ke, Y. Chen, K. Yin, J. Yan, H. Zhang, Z. Liu, J. S. Tse, J. Wu, H. kwang Mao, and B. Chen, Large bandgap of pressurized trilayer graphene, Proceedings of the National Academy of Sciences of the United States of America **116**, 9186 (2019).

[24] M. Wall, The raman spectroscopy of graphene and the determination of layer thickness (2011).

[25] R. P. Dias, C. S. Yoo, M. Kim, and J. S. Tse, Insulator-metal transition of highly compressed carbon disulfide, Physical Review B - Condensed Matter and Materials Physics **84**, 10.1103/PhysRevB.84.144104 (2011).

[26] K. Nagashio, T. Nishimura, K. Kita, and A. Toriumi, Metal/graphene contact as a performance killer of ultra-high mobility graphene analysis of intrinsic mobility and contact resistance (2009) pp. 1–4.

[27] G. Jo, M. Choe, S. Lee, W. Park, Y. H. Kahng, and T. Lee, The application of graphene as electrodes in electrical and optical devices, Nanotechnology **23**, 112001 (2012).

[28] L. Malard, M. Pimenta, G. Dresselhaus, and M. Dresselhaus, Raman spectroscopy in graphene, Physics Reports **473**, 51 (2009).

[29] D. Holec, M. A. Hartmann, F. D. Fischer, F. G. Rammerstorfer, P. H. Mayrhofer, and O. Paris, Curvature-induced excess surface energy of fullerenes: Density functional theory and monte carlo simulations, Phys. Rev. B **81**, 235403 (2010).

[30] A. Bosak, M. Krisch, M. Mohr, J. Maultzsch, and C. Thomsen, Elasticity of single-crystalline graphite: Inelastic x-ray scattering study, Phys. Rev. B **75**, 153408 (2007).

[31] Y. Sun, D. Dunstan, M. Hartmann, and D. Holec, Nanomechanics of carbon nanotubes, PAMM **13**, 7 (2013), https://onlinelibrary.wiley.com/doi/pdf/10.1002/pamm.201310003.

[32] T. Ohta, A. Bostwick, J. L. McChesney, T. Seyller, K. Horn, and E. Rotenberg, Interlayer interaction and electronic screening in multilayer graphene investigated with angle-resolved photoemission spectroscopy, Physical Review Letters **98**, 10.1103/PhysRevLett.98.206802 (2007).
11

# Supplementary Material for: "Pressure induced 3D strain in 2D Graphene"

Nathan Dasenbrock-Gammon[1], Sachith Dissanayake[2] and Ranga P. Dias[1,2,*]

[1]Department of Physics and Astronomy, University of Rochester, Rochester, New York, 14627, USA

[2]Department of Mechanical Engineering, University of Rochester, Rochester, New York, 14627, USA

* Correspondence should be addressed to: rdias@rochester.edu

In this supplementary material, we provide more details of the materials and methods used in this work. The DAC preparation used for this work is intricate and involves many steps, which we provide here:

**Conductivity Setup**

Prior to any graphene transfer, the DAC conductivity setup must be fully prepared. First, diamonds are preindented into a ~250 μm sheet of rhenium to a thickness of ~10 μm to create the gasket. A ~120 μm hole is then drilled into the preindent using Electronic Discharge Machining (EDM). The gasket is then remounted onto one of the diamonds. The top of the rhenium sheet is covered with insulating tape, and then a hole cut around the preindent which is then packed with insulating alumina. The DAC is closed and pressed to form a solid alumina shell. Superglue is applied around the edge of the alumina to adhere it to the rhenium and fill in the gap between the tape and alumina. A hole is drilled into the center of the alumina using a needle to create the sample chamber. Care is taken to ensure the entirety of the rhenium gasket is covered by the insulating alumina. Then platinum leads are cut out of a ~4 μm thick platinum foil and placed into the sample chamber. They are placed such that they will line up with the soon to be sputtered electrodes on the other diamond. A photograph of a completed setup is shown in figure S1.

**Electrode Sputtering**

Once a conductivity setup has been completed atop one of the diamonds, we return to the other diamond. Tantalum electrodes are sputtered onto the diamond using a PVD-75 Lesker Sputter / Evap machine. A vacuum of less than 5E-6 Torr was maintained before beginning the sputtering, which occurred in a 3E-3 Torr argon environment. A custom built shadow mask was used to sputter the electrode shapes. This mask was built from a separate preindented gasket into which the electrode shapes were cut out using an EDM machine with a wedge shaped electrode (photograph of a mask is shown in figure S2). Using a preindented gasket for the shadow mask allows for easy attachment and removal of the mask from the diamond by standard DAC techniques. For some runs a very thin (several nanometers) layer of chromium was first e beam deposited onto the diamond to adhesion of the sputtered electrodes. This was done with the shadow mask in place, and without breaking vacuum on the machine. A thin layer of chromium improves adhesion of the tantalum electrodes to the diamond. An example of electrodes produced by this process are shown in figure S3. After a conductivity setup is prepared and tantalum electrodes sputtered, the DAC is ready for graphene transfer, as described in the main work.

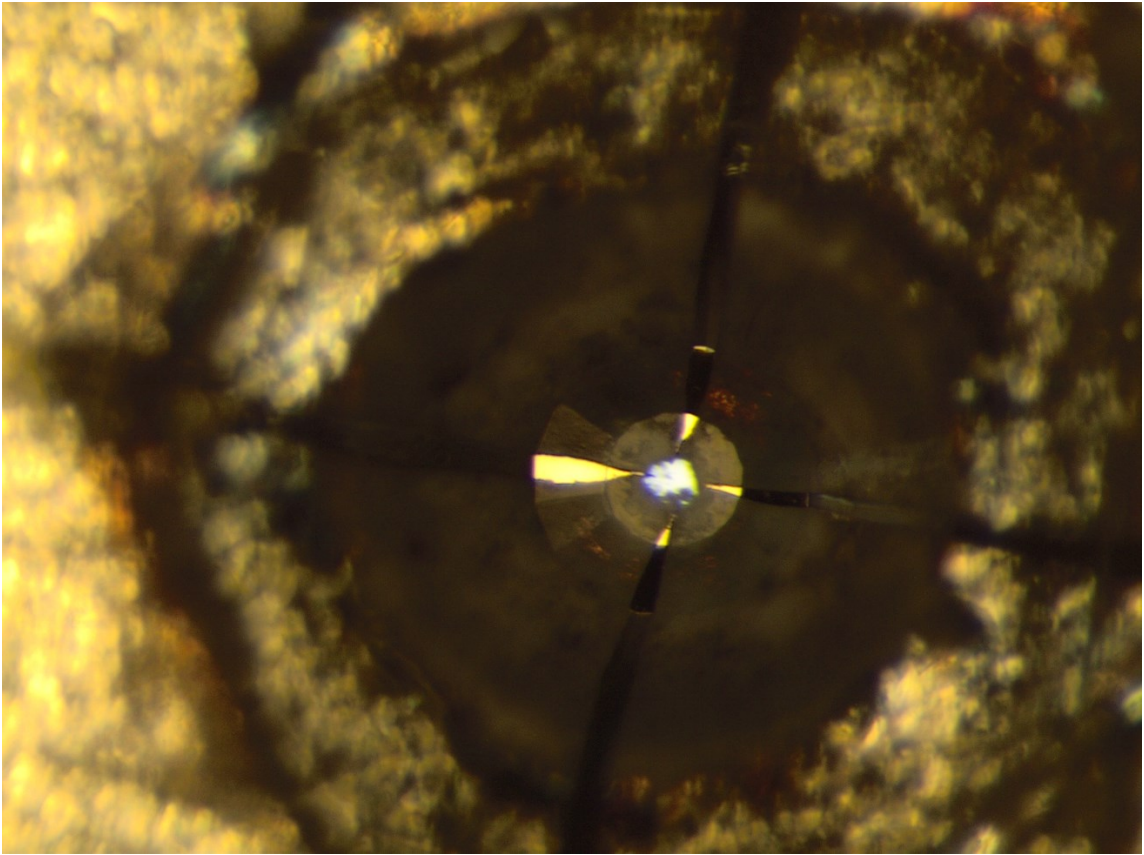

Figure S1: A completed conductivity setup atop one of the DAC diamonds.

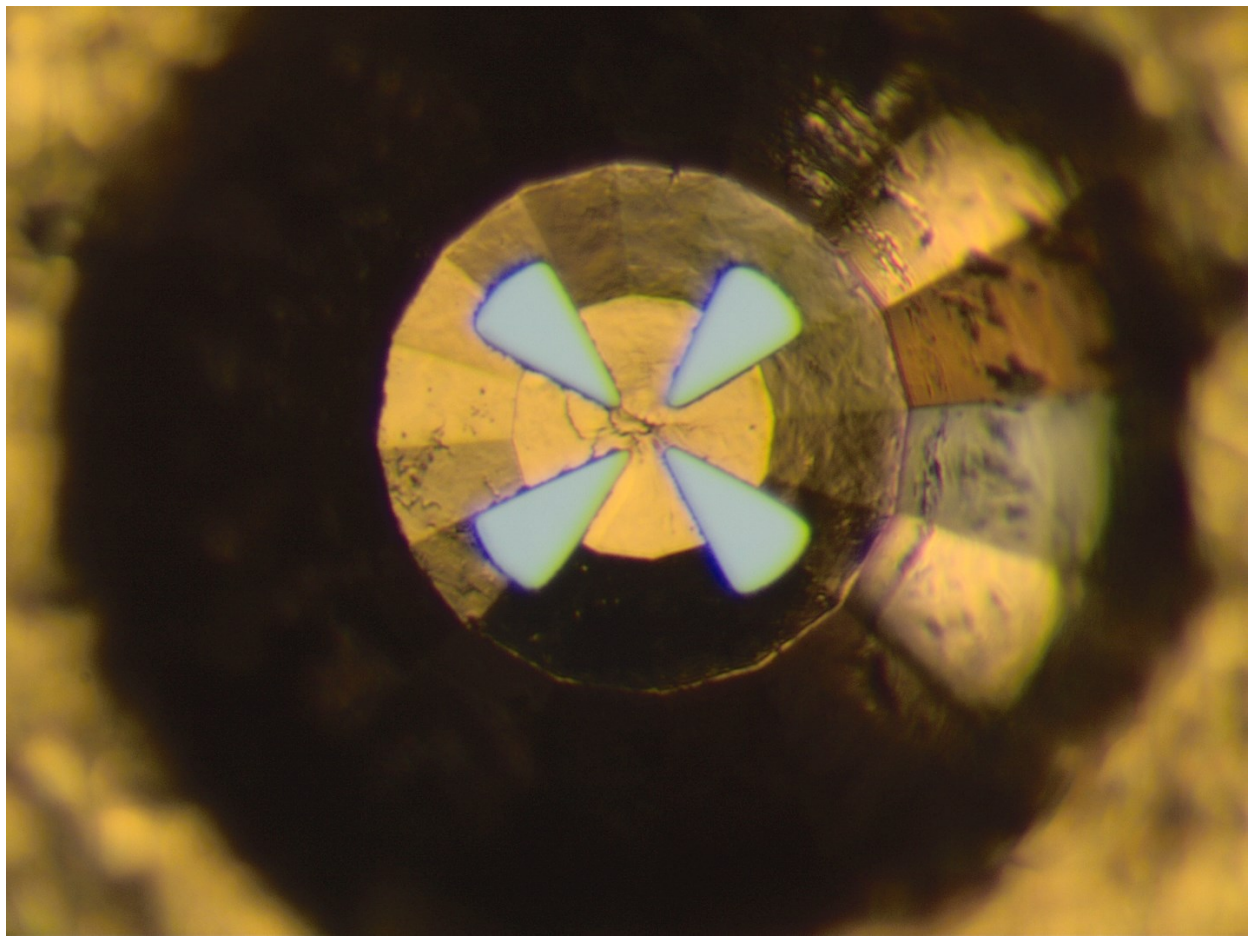

Figure S2: A shadow mask used to sputter the tantalum electrodes onto one of the DAC diamonds.

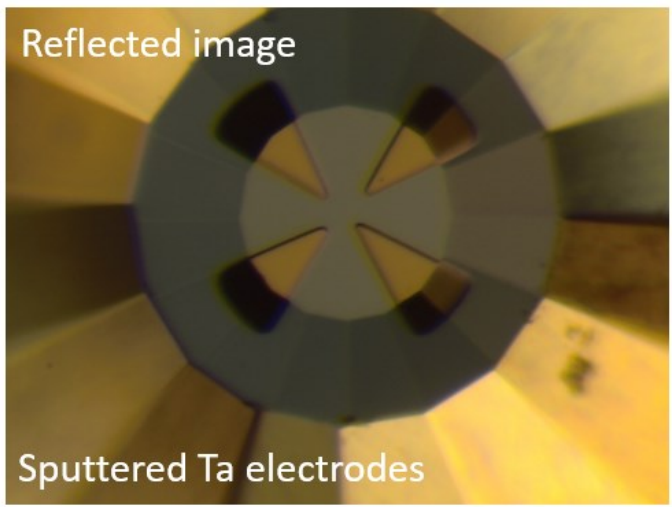 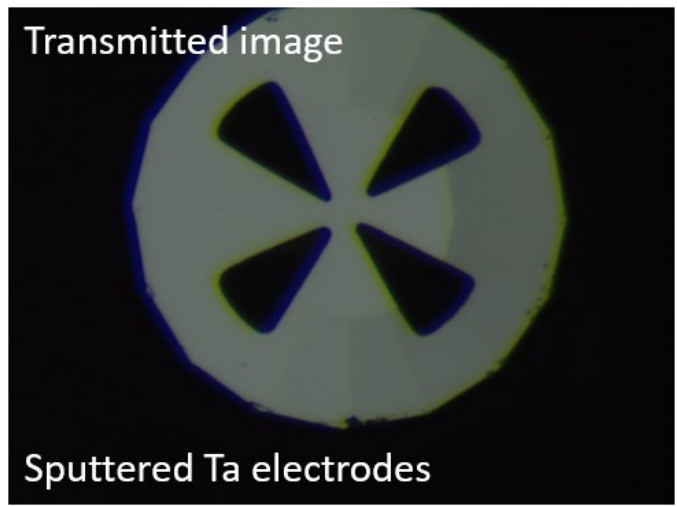

Figure S3: Reflected and transmitted light photographs of sputtered tantalum electrodes.